# Privacy-Aware, Public-Aligned: Embedding Risk Detection and Public Values into Scalable Clinical Text De-Identification for Trusted Research Environments

**Authors:**
Arlene Casey[1]
Stuart Dunbar[1]
Franz Gruber[1]
Samuel McInerney[1]
Matúš Falis[1]
Pamela Linksted[1]
Katie Wilde[2]
Kathy Harrison[1]
Alison Hamilton[3]
Christian Cole[4]

**Affiliations:**
1. Usher Institute, University of Edinburgh
2. Aberdeen Centre for Health Data Science, University of Aberdeen
3. Research and Development NHS Glasgow & Greater Clyde
4. Population Health & Genomics, School of Medicine, University of Dundee; Health Informatics Centre, University of Dundee

**Contact Author:** Arlene.Casey@ed.ac.uk

## Abstract

Clinical free-text data offers immense potential to improve population health research such as richer phenotyping, symptom tracking, and contextual understanding of patient care. However, these data present significant privacy risks due to the presence of directly or indirectly identifying information embedded in unstructured narratives. While numerous de-identification tools have been developed, few have been tested on real-world, heterogeneous datasets at scale or assessed for governance readiness. In this paper, we synthesise our findings from previous studies examining the privacy-risk landscape across multiple document types and NHS data providers in Scotland. We characterise how direct and indirect identifiers vary by record type, clinical setting, and data flow, and show how changes in documentation practice can degrade model performance over time. Through public engagement, we explore societal expectations around the safe use of clinical free text and reflect these in the design of a prototype privacy-risk management tool to support transparent, auditable decision-making. Our findings highlight that privacy risk is context-dependent and cumulative, underscoring the need for adaptable, hybrid de-identification approaches that combine rule-based precision with contextual understanding. We offer a comprehensive view of the challenges and opportunities for safe, scalable reuse of clinical free-text within Trusted Research Environments and beyond, grounded in both technical evidence and public perspectives on responsible data use.

## Keywords:

# Introduction

Clinical free-text offers rich, nuanced information often absent from structured health data, capturing symptoms, disease progression, care decisions, and patient context. Unlocking this unstructured data for research holds enormous potential for advancing biomedical discovery, improving disease understanding, and ultimately transforming patient care. Yet, clinical free text remains one of the least accessible types of electronic health record (EHR) data, primarily due to privacy concerns.

Despite the development of numerous algorithmic de-identification methods[1,2], their real-world application remains limited. Most existing tools are trained and evaluated on cleaned datasets from limited sources, and face challenges in scaling across the diversity and complexity of real-world clinical documentation. In the UK, building a health data infrastructure that is safe, equitable, and research-enabling is a national priority.[3] Achieving this, however, will not be possible without addressing the systematic inaccessibility of clinical free-text, a critical yet underutilised component of the health data landscape.

Realising the full value of free text at scale requires not only technical solutions, but also governance mechanisms that uphold patient confidentiality, ensure transparency, and align with public expectations. Our public engagement findings reinforce that societal support for data use is contingent on meaningful oversight and visible safeguards, not only that privacy is protected, but that the process is explainable and trustworthy.[4,5] Embedding public values into system design is therefore not optional, but essential to enabling ethical data access.

To do this, we must first understand how privacy risks emerge in practice. This includes evaluating how identifiers - both direct and indirect (implicit) - vary across settings, how documentation practices shape risk, and where current de-identification methods fall short. Generating this evidence is essential for informing proportionate and interoperable governance frameworks that reflect real-world clinical and data workflows.

This paper brings together multiple strands of our research on de-identification: empirical analysis of identifier variation across NHS regions and record types; exploration of indirect privacy risks; insights from public engagement on trust, transparency, and acceptability; and the development of a prototype privacy-risk management tool. By synthesising these findings, we offer a comprehensive view of the challenges and opportunities for safe, scalable reuse of clinical free text, and propose early foundations for a framework that supports transparent, governance-aligned NLP in Trusted Research Environments (TREs) and beyond.

# Background and Related Work

Privacy risk in unstructured clinical text has long been a concern in health informatics, driving sustained efforts to develop automated de-identification tools.[1,2] Early approaches relied on rule-based and dictionary-driven systems, while more recent work has explored machine learning and transformer-based language models[6,7], which promise improved generalisability across domains. While these advances show promise in benchmark accuracy, their translation to real-world, operational healthcare environments remains limited, especially where text structure and clinical practice vary widely. Notably, within the UK there has been large-scale release of clinical free-text for research recently.[8] Absent though is transparency on exactly how the de-identification has been carried out although they do only mention de-identification of direct identifiers.

***Challenges in Defining Privacy Risk in Clinical Free-text:*** Unlike structured health data, where identifiers follow standardised formats, privacy risk in clinical free text is highly context-dependent. Sensitive information is often embedded implicitly, with meaning derived from surrounding narrative, document type, or linked data. The same phrase, such as a location or relationship, may carry different disclosure risks depending on context. Indirect identifiers, including references to rare diseases, mental health, social circumstances, or family members, are particularly difficult to detect and assess, especially when their re-identification potential emerges only in combination with additional records.

Regulatory frameworks such as HIPAA[9] and GDPR[10] provide definitions for direct identifiers, but they differ due to jurisdiction e.g. social security numbers in the USA compared to UK national insurance numbers. Additionally, they fall short in covering the subtleties of indirect or contextual risks. Stubbs et al. noted the challenge of annotating indirect risks when preparing clinical data for public release and opted to exclude them, highlighting their complexity and the current lack of annotation standards.[11] Even among expert annotators, disagreement is common regarding what constitutes a privacy-risk entity, introducing subjectivity into gold-standard datasets.[12] Framing privacy risk as a binary classification task therefore oversimplifies what is inherently a layered, context-sensitive process. A more robust, real-world approach must incorporate clinical, technical, and governance perspectives to ensure privacy protection without undermining data utility.

***Gaps in Real-World Evaluation and Generalisability:*** Despite a growing body of research, de-identification models - and clinical NLP models in general - are largely trained and validated on a small set of US based public corpora, most notably i2b2[13] and MIMIC[14]. These datasets, while foundational, are limited in scope, often reflecting documentation from a single institution or limited document types. Moreover, they are sanitised for public release, likely removing many of the indirect (implicit) risks that complicate de-identification in real-world data. This data limitation stifles progress[12] and does not support the need to build tools that can generalise to population level systems.[15]

Flamholz et al. showed that embeddings trained on clinical text can consistently outperform those based on general-domain embeddings for PHI identification.[16] However, other work has shown that pre-trained embeddings do not always translate to better performance, particularly where PHI focus on numeric (e.g. dates, medical numbers) or non-context dependant PHI.[12] Additionally, Johnson et al. show use of BERT based models and due to the sub-word tokenisation, result in a loss of document structure headers which they believe are critical indicators for PHI.[12] Some of the inconsistent reporting of performance on what are the best embeddings, domain or non-domain specific, is likely due to the varying nature of clinical reports used in studies. For example, discharge summaries contain far more varied and contextual PHI than radiology reports. Hybrid models are consistently shown to outperform single model solutions[12,14,17,18] and this is likely due to their ability to address both the contextual and non-contextual nature of PHI.

Without comparative, multi-source studies using actual NHS or regional health datasets, it is difficult to assess whether de-identification tools are genuinely fit for purpose or merely tuned for the conditions of the benchmark. The absence of population-level access to diverse clinical free-text severely limits the field's ability to evaluate generalisability, and in turn, hinders the development of scalable, trustworthy infrastructure for clinical text reuse in research.[12]

***Public Perspectives and the Role of Trust:*** Importantly, public attitudes do not reflect blanket opposition to the use of clinical free text in research. Previous works by Ford et al. highlight that public acceptance is conditional on transparency, ethical oversight, and demonstrable public benefit.[19,20] Participants in public engagement activities have consistently expressed that free text can be used for research but only if safeguards are robust, decision-making is auditable, and individuals' privacy is respected in spirit, not just in regulation.

This reflects a broader shift: privacy-risk management is not solely a technical task but a social one. Tools and frameworks must reflect public values, support transparency, and enable institutions to justify decisions in ways that are understandable and trustworthy. The governance challenges associated with free text - especially around indirect identifiers and contextual disclosure - cannot be resolved by models alone. Addressing these challenges requires systems that combine technical robustness with ethical accountability.

## Methods

We describe briefly the methods used in our existing work, for more details the original papers are referenced.

*Direct Identifiers*
Our aim was to build a labelling schema agreed across the five Scottish TREs and label a set of clinical texts across three TREs on two record types with disparate privacy risks.

*Annotation schema development*: Through a workshop with data analysts, governance, and TRE leads, current approaches to de-identification were discussed. Included in the workshop was a review of existing work in defining direct Personal Health Identifiers (PHIs). The outcome for the workshop was an agreed annotation schema for labelling direct PHIs. Further details of annotation labels, arrangements can be found here in Casey et al.[21]

*Direct Identifier Data:* 2000 Discharge Summaries and 2000 X-ray Radiology reports which represent different health record types. The Discharge Summaries were selected based on the main hospitals within the regional TRE for adult patients (18+) who were admitted to a ward with at least a 24-hour stay in 2022. Random selection of patients was undertaken stratifying across age bands, 18-30 years, 31-40 years, 41-50 years, 51-70 years, 71 years and over.

*Indirect or Implicit Identifiers*

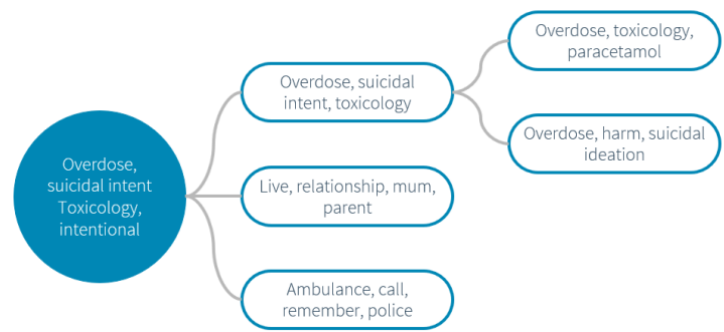

*Figure 1- Example sentence-based Clusters from BERT Topic Modelling*

Our aim here was a scoping exercise to better understand what indirect identifiers were and how they occur. These types of identifiers are rare, very contextual and nuanced. Our approach was to use sentence similarity with hierarchical clustering to explore patterns that expose indirect risks.[22,23] We pre-processed reports into sentences and applied BERT Topic modelling[24] at a sentence level, using similarity measures to cluster sentences together. We labelled the topics using class-based TF-IDF (see Figure 1, an example of sentence clusters and labels). These topic labels were manually reviewed for any that may contain privacy risks and for those that contained privacy risks the reports were read. This qualitative approach allowed us to direct our attention to records containing indirect risks. Results were reviewed with Information Governance and clinical staff.

*Indirect Identifier Data*: One year of hospital Discharge Summaries (2022), from three major hospitals in NHS Lothian, for patients with admission to a hospital ward with a >24 hours stay; this resulted in approximately 89,000 reports. Data was split across age ranges of patients, 18-30 years old, 31-50 years old, 51-70 years old and 71 years and over.

*Public Perspectives on clinical free text and NLP to de-identify free text*

We conducted a series of public engagement activities, independently facilitated and analysed by IPSOS Scotland[25], to explore perceptions and concerns surrounding the use of clinical free text in health data research. These activities included a survey and workshops consisting of an information session followed by a deliberative workshop. More details can be found in the independent IPSOS report and our SARA PPIE report.[26,27] The sessions were designed to facilitate open discussions about the nature of clinical free-text data, its potential benefits for research, and the associated privacy considerations. Participants were provided with information on how clinical free text is generated, the types of information it may contain, and the methods used for de-identification. Feedback was collected through qualitative methods, including thematic analysis of discussion transcripts and participant feedback forms. This captured nuanced views and informed the development of governance process for clinical free-text data and our approaches to de-identification, in addition to the design of a prototype privacy-risk management and audit tool.[28]

*Development of a Privacy Risk Tool*

The aim was to develop a prototype dashboard to visualise privacy risks in clinical free text, to support Information Governance and data analysts within TREs during the data ingress process for a research project. The goal was to enable proportionate, risk-based decisions by providing cohort-level insights into potential privacy concerns.

Initial mock-ups were co-designed with stakeholders across the Scottish TRE network and facilitated by a specialist design company. While the current version focuses on visualising and exploring risks, the full design workflow included future functionality for tracking and auditing de-risking decisions.

## Results

*Direct Identifiers*

The frequency of entities varied significantly by document type, with discharge summaries generally containing between two to three more times the number of identifiers than Radiology reports, although this varies between TRE and identifier type. We observed a wide discrepancy between the number of identifiers across the TREs. Note TRE 3 only reported counts for 20% records, so their total number of identifiers is estimated in the final column. Names and titles were frequent but less in radiology reports, and many of these tended to be clinician names in Radiology reports. TRE 3 had a far higher number of names or contact identifiers than the other 3 TREs but interestingly they had far less references to a patient's age. Dates were a common feature across both record types. All TREs had relatively few references to ethnicity across both record types. IDs, particularly CHIs were much higher in TRE3.

*Table 1- Annotated Entity Counts for each report types and each TRE – TRE 3 is extrapolated to match 2000 records from 20% counts; Contact Entities are bundled together*

|  | TRE 1 | TRE 2 | TRE 3 | TRE 1 | TRE 2 | TRE 3 |
|---|---|---|---|---|---|---|
|  | Discharge Summaries | | | Radiology Reports | | |
| **Names (First & Last)** | 10,726 | 6,171 | 31,050 | 4,380 | 1,699 | 12,940 |
| **Title** | 5,512 | 3,497 | 13,755 | 1,892 | 52 | 5,640 |
| **Initials** | 1,643 | 388 | 7,020 | 173 | 74 | 15 |
| **Age** | 334 | 556 | 290 | 470 | 46 | 65 |
| **Occupation** | 6,748 | 2,794 | 4,335 | 2,251 | 924 | 5,495 |
| **Ethnicity** | 2 | 42 | - | - | - | 5 |
| **Hospital** | 3,079 | 1,170 | 4,675 | 106 | 7 | 1,480 |
| **Ward** | 3,739 | 21,95 | 995 | 38 | 25 | - |
| **Dates** | 5,578 | 960 | 13,660 | 3,707 | 347 | 7,870 |
| **Contact (Address, Postcode, Town, Country, Phone, Org & building name)** | 1,910 | 294 | 42,485 | 117 | 48 | 2,850 |
| **IDs (e.g. CHI, GMC)** | 104 | 29 | 5,935 | 4,388 | 516 | 2,260 |

*Table 2 - Annotator F1 Scores, over all entities, each TRE (1-3) and each report type (Discharge Summary, Radiology)*

| EHR Type | TRE | First Name | Last Name | Title | Initials | Age | Dates | Year | Hospital | Ward | Occ-upation |
|---|---|---|---|---|---|---|---|---|---|---|---|
| **Discharge Summary** | TRE 1 | 95 | 95 | 99 | 72 | 94 | 97 | 95 | 95 | 93 | 79 |
|  | TRE 2 | 88 | 91 | 93 | 80 | 69 | 74 | 72 | 89 | 64 | 69 |
|  | TRE 3 | 60 | 70 | 71 | 35 | 38 | 48 | 25 | 53 | 52 | 63 |
| **Radiology Report** | TRE 1 | 99 | 95 | 99 | 77 | 94 | 99 | 95 | 94 | 67 | 97 |
|  | TRE 2 | 96 | 95 | 98 | 87 | 77 | 81 | 80 | 71 | 49 | 68 |
|  | TRE 3 | 97 | 98 | 99 | 67 | 92 | 96 | 32 | 91 | - | 97 |
| EHR Type | TRE | Address Line | Town | Country | Post Code | Phone | Org Name | Building Name | GMC | CHI | UHPI |
| **Discharge Summary** | TRE 1 | 80 | 63 | 67 | 92 | 83 | 09 | 0 | 100 | 100 | 60 |
|  | TRE 2 | 15 | 77 | 64 | 92 | 50 | 45 | 64 | 08 | 71 | 0 |
|  | TRE 3 | 13 | 66 | 80 | 71 | 53 | 63 | 17 | 0 | 68 | 0 |
|  | TRE 1 | 0 | 100 | 67 | - | 100 | 67 | - | 98 | - | - |
| Radiology Report | TRE 2 | 100 | 40 | 100 | 100 | 100 | 62 | - | 54 | 98 | - |
| | TRE 3 | - | - | 92 | - | 98 | 96 | - | 99 | - | - |

Among the TREs Radiology reports human agreement of F1 scores is much higher than Discharge Summaries. There are fewer entities within Radiology reports. Scores were generally higher in TRE 1, which could be attributed to using pre-annotation rule-based labelling for structured items (Dates, Post Code, Hospitals etc). TRE 2 had almost 50% fewer names or address-related information compared to TRE 1. This is a technical artifact of how data are processed, which allows TRE 2 to automatically remove many entities that would normally be placed within headers or footers in these EHRs. Discharge Summaries revealed greater variability, especially for semi-structured fields like Address Line and Ward. Notably, entities like Initials demonstrated inconsistent annotation quality across document types, reflecting challenges in disambiguating short or infrequent tokens and seen in other works.[29]

Patterns of disagreement around entity types and boundaries were found such as: Double first or last names, marking of dates and years inconsistently, and boundaries on age (often written as 46yo or 46F). Local hospitals marked as two entities i.e. town and a hospital (e.g. Fife hospital) particularly in one TRE which supported more rural areas. Phone numbers where they included extensions. Wards and clinics can be hard to recognise when named after clinicians or community places which caused disagreement on entity type or boundaries. Occupations mentions, for example, 'the police found the patient', disagreement occurred as to whether this occupation mattered in the context of privacy risk.

Additionally, we carried out a manual review of an existing proprietary de-identification system output against the annotated data, looking specifically at Common Health Index (CHI) and dates. Comparing the third party and human labelled data the performance had dropped to an F1 score of 80% for both these entities compared to an implementation target of >95%.

*Implicit (Indirect) Identifiers*

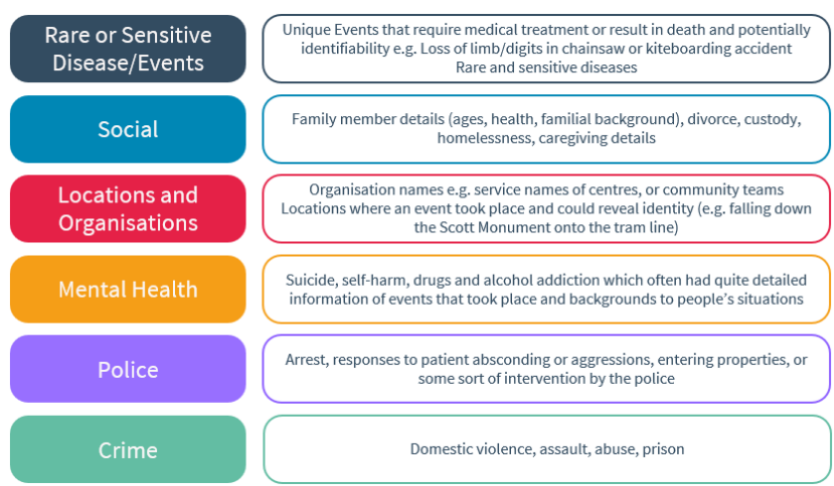

Figure 2 - Indirect (Implicit) Identifier Categories

Analysis revealed six broad categories that indirect (implicit) risks fall under (Figure 2): Unique Medical Diseases or Events, Social, Locations & Organisations, Mental Health, Police, and Crime. The occurrence of these risks was not isolated; often, where one category surfaced within a sentence, a cascade of others could be found in subsequent reports. The age range of patients and the frequency of attendance emerged as crucial factors influencing the cumulative risk of identifiability.

*Privacy Risk Tool*

The prototype was developed using R Shiny to align with tools already widely accepted across the TRE eco-system and avoid additional governance hurdles of software permissions. The dashboard enables users to upload a labelled cohort and explore report types, demographics (e.g. sex, age, ethnicity, Scottish Index Multiple Deprivation), and time range (Figure 3). Filters allow dynamic views of risk distributions across these variables. Users can view summary tables of risk types, explore co-occurring risks within reports, and drill down to individual or patient-level risk profiles.

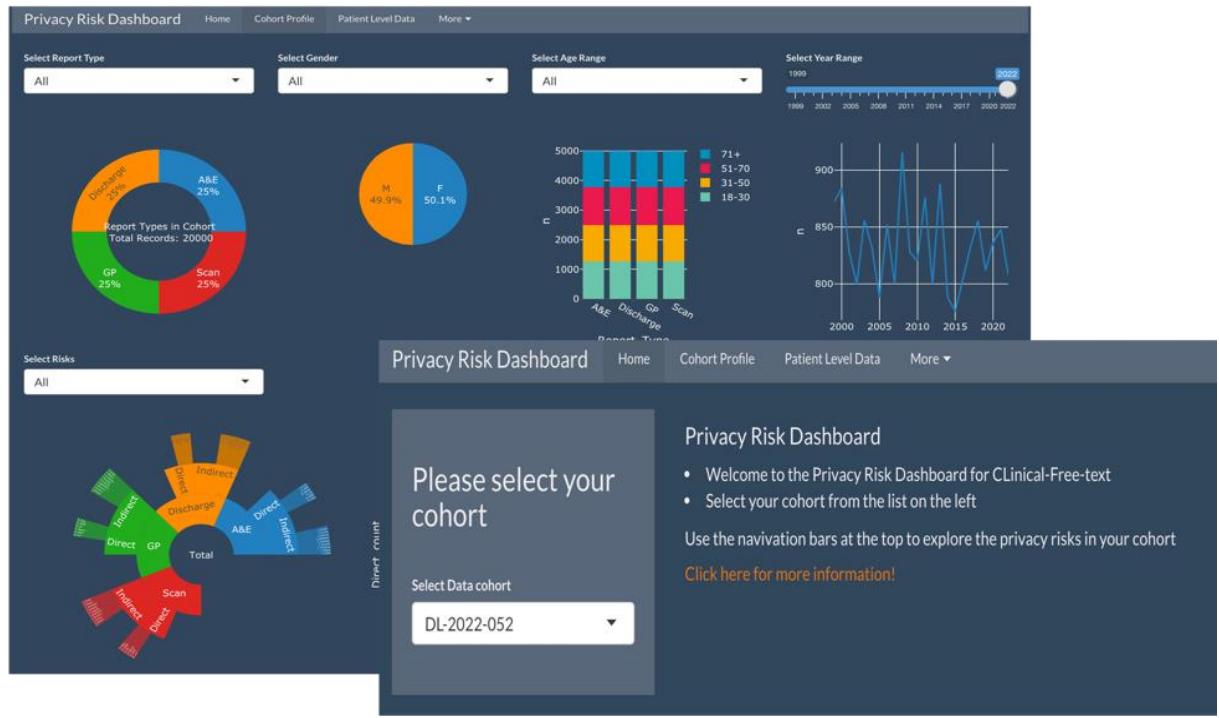

*Figure 3- Example of Privacy-Risk Dashboard tool*

*Public Engagement*

Participants recognised the richness and potential value of clinical free text for improving care, especially for areas like mental health and rare conditions where structured data may fall short. However, they also expressed clear concerns around privacy risks, particularly when narratives include detailed personal histories or contextual clues that could lead to re-identification. There was broad support for research use, provided strong safeguards were in place. Transparent governance, human oversight, and demonstrable public benefit were emphasised as conditions for trust, this was built into the workflow for the privacy-risk tool. Participants valued layered, explainable risk assessments and supported the idea of tools that could make privacy risks visible and manageable within data access decisions.[26,27]

# Discussion

*Direct Identifiers*

Our analysis highlights significant variation in the presence and format of direct identifiers across hospital sites, driven largely by differences in clinical documentation practices, system configurations, and TRE operating models. For example, some hospitals using TrakCare employ templated note-forms that automatically insert metadata such as hospital name, ward location, and clinician detail, resulting in predictable patterns of identifier inclusion. The configuration of the TRE also plays a role: some TREs are embedded within NHS infrastructure with direct access to source systems, while others are externally hosted and receive pre-processed document outputs. This means in contrast, some sites generate clinical letters as PDFs, which are later converted to text using optical character recognition (OCR). These PDFs often contain more embedded patient details due to additions during processing. In other instances, some clinical writing (e.g. radiology report writing) is outsourced, leading to greater uniformity in structure but less consistency in how identifiers appear. Whilst this highlights differences in PHI across regions within a nation it is also worth noting that this happens across the UK. The work in Kraljevic et al., a UK based de-identification tool, highlights that they see a need for annotations and removal of PHI such as passports and driving licenses.[29] This is something that we did not see in our data. On discussion with clinical professionals about their experiences of clinical free text within Scotland this was confirmed. These factors create both challenges and opportunities for de-identification approaches, underscoring the need for adaptable methods that can accommodate local system and workflow differences.

*Annotator feedback* was critical in helping us understand how different entity types present within and across document types, with direct implications for the design of de-identification solutions. Some identifiers, such as postcodes and dates, were typically easier to detect due to their consistent formatting and predictable placement, making them well-suited to rule-based methods. Patient names, by contrast, were rare in radiology reports but more common in discharge summaries, often embedded in narrative text where they are less easily identified through surface patterns alone. Similarly, hospital and ward names were frequently inserted via templates in one TRE but also appeared contextually in free text, causing variation by site and system. Annotators also noted contextual ambiguities, for instance, distinguishing between patient names and family members and hospital names and places, highlighting the limitations of standalone methods and the potential value of hybrid approaches. These challenge of understanding what constitutes PHI is not new[12] but it underlines the need for empirical evidence on privacy risks and open discussion on how to address and agree on these.

*Indirect (Implicit) Identifiers*

Our analysis of indirect privacy risks highlights the critical need to move beyond traditional de-identification approaches that focus solely on direct identifiers. Our six broad categories of indirect risk frequently appeared in overlapping or cascading patterns within and across documents. These findings underscore that implicit disclosures are not edge cases, but recurring elements that carry significant re-identification potential, particularly when considered in combination or in data linkage scenarios. Patient age and frequency of healthcare interactions were also found to increase identifiability, further compounding risk. This has important implications for the development of privacy-risk management systems: treating these categories as secondary or optional risks underestimates their cumulative impact. Instead, models and governance workflows must be designed to systematically detect, log, and evaluate such risks. This calls for integrated approaches that can reason over context and temporality, and tools that support nuanced, cohort-level assessment, ensuring that privacy protection keeps pace with the realities of narrative clinical data.

Together, these findings underscore the need for adaptable de-identification systems that can tailor strategies by document type and identifier behaviour, combining rules, domain-aware embeddings, and contextual reasoning to manage risk effectively across diverse clinical settings.

*Need for Continual Monitoring of System Performance*

The performance drops when comparing outputs from a proprietary de-identification system to our gold-standard annotated labels illustrates a critical challenge in deploying de-identification systems in dynamic, real-world environments. Further analysis revealed that changes in how certain identifiers presented had occurred since the system's original training phase. These shifts, introduced through updates in clinical documentation templates and processing workflows, underscore the importance of ongoing system monitoring and evaluation. Models, even those that initially perform well, can degrade in accuracy over time as data formats and content evolve. This finding reinforces the need for continuous validation and maintenance of de-identification pipelines to ensure they remain robust, particularly in settings where document structure and language are shaped by local operational practices. This places a responsibility on TREs to ensure transparency in how such tools are implemented and monitored. Likewise, third-party tool providers must support deployment in dynamic clinical settings. Given limited financial and technical resources within TREs and the NHS, solutions must be designed with sustainability and adaptability in mind.

*Public Engagement*

Public Engagement played a vital role in shaping our understanding of how clinical free text should be handled in research settings. Participants acknowledged the value of free text for improving care and capturing complexity, particularly in underrepresented areas like mental health and rare conditions. However, they also expressed clear and consistent concerns about re-identification risks, especially where contextual details or narrative patterns could inadvertently reveal a person's identity. Like previous research[19,20], public contributors did not oppose the use of free text but emphasised the need for robust safeguards, clear governance, and meaningful oversight. These findings reinforce the importance of embedding public values into the design of de-identification tools and workflows. They also suggest that technical solutions alone are insufficient. Trust in the use of clinical free text depends on transparency, explainability, and demonstrating that privacy risks are actively assessed and responsibly managed. As such, public engagement should not be treated as peripheral, but as a core component of ethical data science.

*Managing Privacy-Risk*

Our prototype privacy risk management tool was a first step and reflects the strong-demand from the public that there is transparency, traceability and proportionality in privacy-risk decisions. Rather than replacing expert review, the tool is designed to augment existing workflows and help embed structured risk assessment into TRE workflow. As demand for free-text access increases, this tool offers a foundational step toward integrating real-time privacy-risk management into routine data governance.

## Recommendations

Our findings underscore the need for de-identification strategies that are context-aware, flexible, and embedded within wider privacy-risk governance frameworks. While rule-based methods are effective for well-structured identifiers (e.g. postcodes, NHS numbers), more complex, implicit risks, such as references to mental health, rare conditions, or social circumstances, require models that can reason over language and context. Public engagement clearly indicates that these risks are meaningful and must not be treated as secondary.

We recommend the adoption of hybrid de-identification pipelines that combine deterministic and machine learning approaches, tailored by document type and use case. In addition, privacy-risk assessment should not end at model output: it must be auditable, explainable, and accessible to governance teams. Tools like the one developed - providing visualisation, cohort-level summaries, and decision logging - should be adopted to support transparent, risk-informed data access processes within TREs. Lastly, regular validation and review of de-identification systems are essential, particularly in response to evolving clinical documentation practices and system configurations.

## Conclusion

This work contributes a comprehensive, real-world evaluation of privacy risk in clinical free text, demonstrating how direct and indirect identifiers manifest differently across NHS sites, systems, and record types. By integrating NLP methods, empirical annotation, public engagement, and governance tool development, we offer a multi-dimensional perspective on what it takes to enable ethical, scalable access to unstructured health data. Public contributors supported the use of clinical free text in research, but clearly articulated the need for visibility, accountability, and safeguards around privacy. Our prototype privacy-risk management tool offers a step forward in bridging technical and governance domains, supporting risk transparency and cohort-level oversight and we will continue to develop this. Together, these findings offer a foundation for building trusted, adaptable de-identification solutions that are responsive to both operational complexity and societal expectations. As the secondary use of free text expands, embedding public values and real-world evaluation into system design will be essential to maintaining trust and unlocking the full research potential of clinical narratives.